\documentclass[preprint,article,letterpaper]{revtex4}
\usepackage{amsmath}
\usepackage{epsfig}
\usepackage{rotating}
\usepackage{dcolumn}

\newcommand{\str}{$^\mathrm{s}$}

\def\sun{\hbox{M$_\odot$}}

\begin{document}

\author{Benjamin Monreal}

\email{bmonreal@mit.edu}
\affiliation{Massachusetts Institute of Technology\\
Laboratory for Nuclear Science}
\title{Cosmic-ray strangelets in the Earth's atmosphere}

\begin{abstract}
If strange quark matter is stable in small lumps, we expect to find such lumps, called ``strangelets'', on Earth due to a steady flux in cosmic rays.  Following recent astrophysical models, we predict the strangelet flux at the top of the atmosphere, and trace the strangelets' behavior in atmospheric chemistry and circulation.   We show that several strangelet species may have large abundances in the atmosphere; that they should respond favorably to laboratory-scale preconcentration techniques; and that they present promising targets for mass spectroscopy experiments.        
\end{abstract}

\maketitle

\section{Introduction}\label{intro}

A dense ball of up, down, and strange quarks, called a ``strangelet'', may be the true ground state of nuclear matter\cite{witten}\cite{farhi}.  A strangelet is stabilized by its filled quark energy levels; the extra flavor degree of freedom provided by the strange quark lowers the Fermi energy of the up and down quark seas.  The strangelet may be absolutely stable if this decrease in Fermi energy compensates for the extra mass of the strange quark; this may be true above some threshhold baryon number $A_t$.  (Throughout this paper, we will speak as if $A_t$ is ``small'', in the range 50--4000, and therefore that ``small'' strangelets with are stable.  Of course, this is the unknown we wish to study.)

Small stable strangelets have approximately equal numbers of u, d, and s quarks.  A slight deficiency of s quarks gives the strangelet a net positive charge of approximately $Z = 0.3 A^{2/3}$ for the color-flavor locked (CFL) model\cite{madsen:CFL} or $Z = 0.1A$, turning over to $\sim 8A^{(1/3)}$ for large A, for the MIT bag model\cite{farhi}.  (We discuss both models with a strange quark mass of 150 MeV.)  Ordinary nuclei cannot decay into strangelets in the the lifetime of the Universe.  However, a neutron star may quickly convert into a ``strange star'', essentially a large strangelet with $A \sim 10^{50}$\cite{madsen:star}.  Strange star collisions\cite{lee02}\cite{caldwell91} may release smaller strangelets with a range of sizes, including low-mass strangelets with charges $Z$ where $Z_{min} \leq Z  < 100$ and baryon number $A$ where $A_t \leq A < 10^5$.   These strangelets are accelerated by astrophysical shocks and propagate throughout the Galaxy.  Therefore, if small strangelets are stable at all, we expect to find them among energetic cosmic rays\cite{madsen:strangelets} and in cosmic-ray targets like the Earth and Moon.  (Although this same cosmic-ray flux would also contribute strangelets to Earth's progenitor material---the presolar nebula, etc.---we neglect this contribution, and aim for a conservative estimate based on the flux over the Earth's history.)

At rest, a strangelet of charge $Z$ behaves like a heavy isotope of atomic number $Z$; it has the same number of electrons, chemical and thermodynamic behavior, etc..  Thus, we give a ``Periodic Table of Strangelets'' in Figure~\ref{pertab}.   Its chemistry is modified by isotope mass effects which are discussed in section~\ref{secDist} and considered to be small; we will neglect these effects for all discussions of geology and atmospheric chemistry.   

Strangelets can be studied by a mass-spectroscopic search for rare, ultra-heavy isotopes of ordinary elements. In order to search for strangelets with high sensitivity, we must find samples where the strangelet concentration is as high as possible.  Because we can purify a sample chemically before doing mass spectroscopy, strangelets of charge Z will be mixed---or diluted, or contaminated---only with nuclei of charge Z.   The chemical abundances of strangelets are uncorrelated with the abundances of ordinary elements on Earth; therefore, it is helpful to search for strangelets among rare elements.  To illustrate, calcium (Z=20) is abundant and scandium (Z=21) is rare, on Earth, due to details of nuclear structure and nucleosynthesis.  We have no reason to expect that ``strange scandium'' (\str Sc) is any less abundant than \str Ca.   Broadly speaking, then, a search for \str Sc in a Sc sample is more likely to find high concentrations than a search for \str Ca in a Ca sample.  

Several searches have been done along these lines, most commonly in meteorites, Earth rock, and moon rock\cite{hemmick}\cite{Klingenberg}.  However, rock of any sort is a very dirty environment: arriving strangelets are immediately mixed with large numbers of ordinary nuclei of all charges\cite{feiveson}.   Before strangelets reach the Earth's crust, they stop in a very clean environment: the atmosphere.  Studying the chemical and circulation properties of the atmosphere suggests new places to search for small stable strangelets.  Surprisingly high concentrations can be found in the atmosphere itself\cite{mueller}.  In this paper, we suggest a series of atmospheric strangelet searches which together cover a wide range of allowable strangelet charges and masses, in some cases making it possible to probe astrophysical models using existing mass spectroscopy techniques. 

\subsection{Expected strangelet flux}

We derive abundance estimates from the detailed cosmic-ray propagation model of \cite{madsen:strangelets}.  This work assumes a strangelet production rate of $10^{-10}$ solar masses (\sun) per year in our Galaxy, with $10^{-5}$\sun\ of strange matter released per collision, and a collision every 30,000 y.  It includes the effects of acceleration, interstellar propagation, and solar modulation using several phenomenological models.  If we assume that all of the released strange-star material is converted into strangelets of charge Z and and mass A (GeV/c$^2$), we derive the maximum flux F

\begin{equation}\label{eqFlux}
F = 2 \times 10^5 \mathrm{m}^{-2} \mathrm{y}^{-1} \mathrm{ster}\times A^{-0.467} Z^{-1.2} R_{\mathrm{cutoff}}^{-1.2}
\end{equation}

where $F$ is the flux in particles m$^{-2}$ ster$^{-1}$ y$^{-1}$, and $R_{\mathrm{cutoff}}$ is the either the geomagnetic cutoff or the solar-modulation cutoff, whichever is greater.  Notice that, although a factor of $A^{-1}$ is expected when a finite pool of strange matter is partitioned into strangelets, heavier strangelets propagate more efficiently, giving the shallow A dependence.  Equation~\ref{eqFlux} gives an upper limit on the flux because it assumes that \emph{all} of the produced strangelets have the same charge.   More realistically, we expect strange star collisions to produce a wide range of strangelet charges, thereby lowering the number of strangelets with any particular charge. 

\begin{figure}
\epsfxsize=4.5in
\begin{center}
\epsfbox{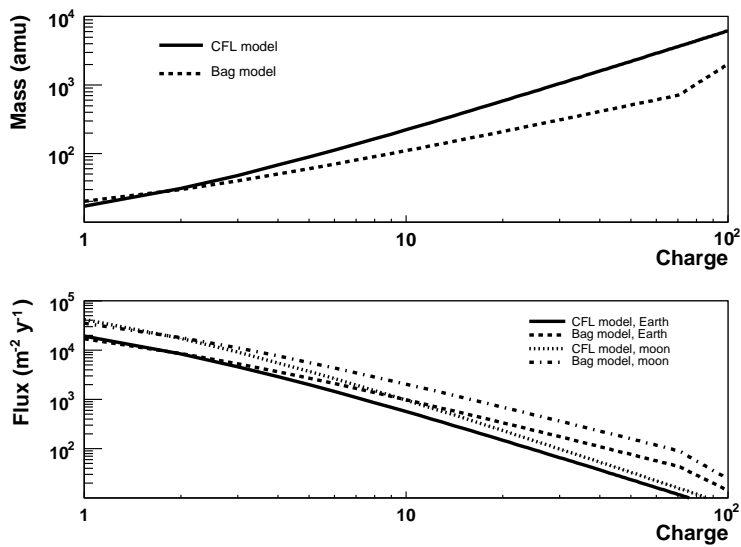}
\end{center}
\caption{Top: Mass versus charge for CFL strangelets (solid) and bag-model strangelets (dashed).  Bottom: Expected cosmic-ray strangelet flux versus charge, from Equation~\ref{eqFlux}, in particles m$^{-2}$ y$^{-1}$ averaged over the planet.  The solid (dashed) line shows CFL (bag-model) strangelets at Earth, including both solar modulation and the geomagnetic cutoff.  The dotted (dash-dotted) line shows CFL (bag-model) strangelets at the moon, with solar modulation but no geomagnetic cutoff. }\label{figF}
\end{figure}

We compute the average flux at Earth by integrating {Equation~\ref{eqFlux}} over the Stormer dipole approximation for the geomagnetic cutoff\cite{stormer2}, and approximating the solar modulation cutoff as $R = (A/Z)^{1/2}(\Phi/500 MV)^{1/2}$\cite{madsen:strangelets}, where the solar modulation parameter $\Phi$ is taken to be 500 MV.  These fluxes are given in Table~\ref{TableNG} and Figure~\ref{figF}.  

\section{Strangelets on the Earth, moon, and meteorites}

As a basis for comparison, we review the expected abundances of strangelets in the Earth's crust, on the moon, and in meteorite material. 

The moon\cite{taylor75} is the simplest case.  Most of the moon's surface, called the ``highlands'', has been essentially stationary for about $\sim$4 billion years.  Its surface has been turned over and reburied to a depth of $\sim$10 meters by the action of meteor impacts.  It has no substantial magnetic field.  If conservatively assume that the surface turnover/reburial rate is constant over time, then the expected strangelet per-atom concentration in the lunar highlands is \mbox{$c_{\mathrm{moon}} = F \times 3 \times 10^{-20}$}, where $F$ is the Earth-averaged annual flux per m$^2$ y from equation~\ref{eqFlux}.  For example, for \str O (Z=8, A=130) the concentration is $2\times10^{-16}$.  The turnover rate has been slower in the modern epoch; the top 10cm of soil may have been exposed, at many sites, for 500 My\cite{eberhart}, leading to a proportionally higher concentration.

Meteorites are more complex as strangelet targets; cosmic-ray exposure measurements\cite{begemann69}\cite{anders62} of most meteorite samples show quite short ages $T_{CRE}$, often of order 10 My, suggesting that the sampled material spent most of its life ``buried''---and shielded from cosmic ray exposure---inside a larger object.  Furthermore, much of this exposure is due to cosmic-ray protons, which penetrate much more deeply than high-Z strangelets.   If we ignore these and other complications\footnote{For example, a meteoroid in the asteroid belt would encounter a lower solar modulation cutoff.}, then the strangelet abundance is roughly proportional to the cosmic-ray exposure age, \mbox{$c_{\mathrm{meteor}} = F \times 10^{-23} \times \frac{T_{CRE}}{1 My}$}.   

The Earth's crust and oceans are a particularly complex environment, some details of which we will address in a forthcoming paper.  Over the Earth's 4 billion year history, strangelets have come out of the atmosphere and been deposited on land and in the ocean.  Initially, strangelets are probably incorporated into sedimentary rock.  This rock can be buried, subducted, metamorphosed, uplifted, and eroded by various processes, while new (possibly strangelet-free) rock wells up from the mantle below.  For a rough estimate, we may neglect the details and suppose that the Earth's strangelet inventory is evenly mixed through all sedimentary rocks.  The average depth of sedimentary material is about 2 km over the Earth's surface, and this layer has re-cycled with the mantle about 5 times\cite{feiveson}\cite{mohrig}.  This suggests that Earth's integrated strangelet flux is evenly mixed with a 10km-deep column of rock, where the strangelet abundance is approximately $c_{\mathrm{crust}} = F \times 1.5 \times 10^{-23}$.  

\section{Strangelets in the atmosphere}
\begin{sidewaysfigure*}
\epsfxsize=8.0in
\begin{center}\epsfbox{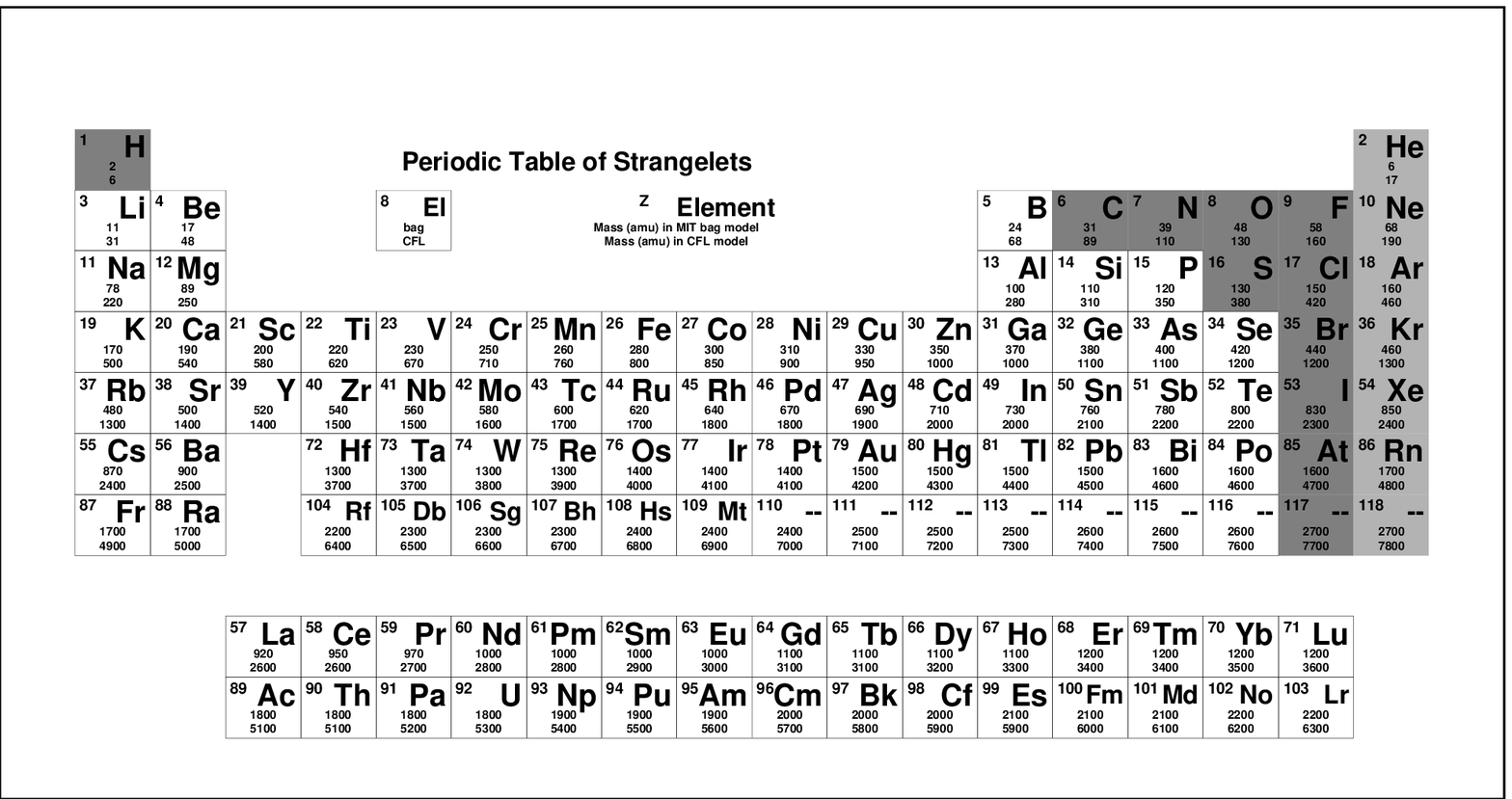}\end{center}
\caption{The Periodic Table for strange quark matter, showing the predicted mass range.  Noble gases are shown in light grey, ``volatiles'' are in dark grey.  The remainder are considered ``metallic'' in terms of their atmospheric behavior.  The strangelet masses are calculated both for the CFL charge-mass relation $Z=0.3 A^{2/3}$ and the MIT bag model relation $0.1A \leq Z \leq 8A^{1/3}$}\label{pertab}
\end{sidewaysfigure*}

A cosmic ray strangelet has less penetrating power than a nucleus of the same $Z$ and energy\footnote{Stopping distances can be calculated with SRIM, \copyright 2000 by James Zeigler, http://www.srim.org} due to a lower velocity and higher dE/dx.  Therefore, most strangelets will lose energy by ionization and stop the mesosphere and upper stratosphere, 50--80 km above the surface.  After stopping, their fate is determined by atmospheric chemistry and dynamics.  We will consider three broad classes of chemical species the atmosphere: noble gases, volatile strangelets, and metallic strangelets.  These categories are defined in Figure~\ref{pertab}. 

\subsection{Effects of atmospheric circulation}

Several details of atmospheric physics\cite{salby} are particularly relevant to the behavior of metallic strangelets.

\begin{figure*}
\begin{center}
\epsfxsize=6.0in
\epsfbox{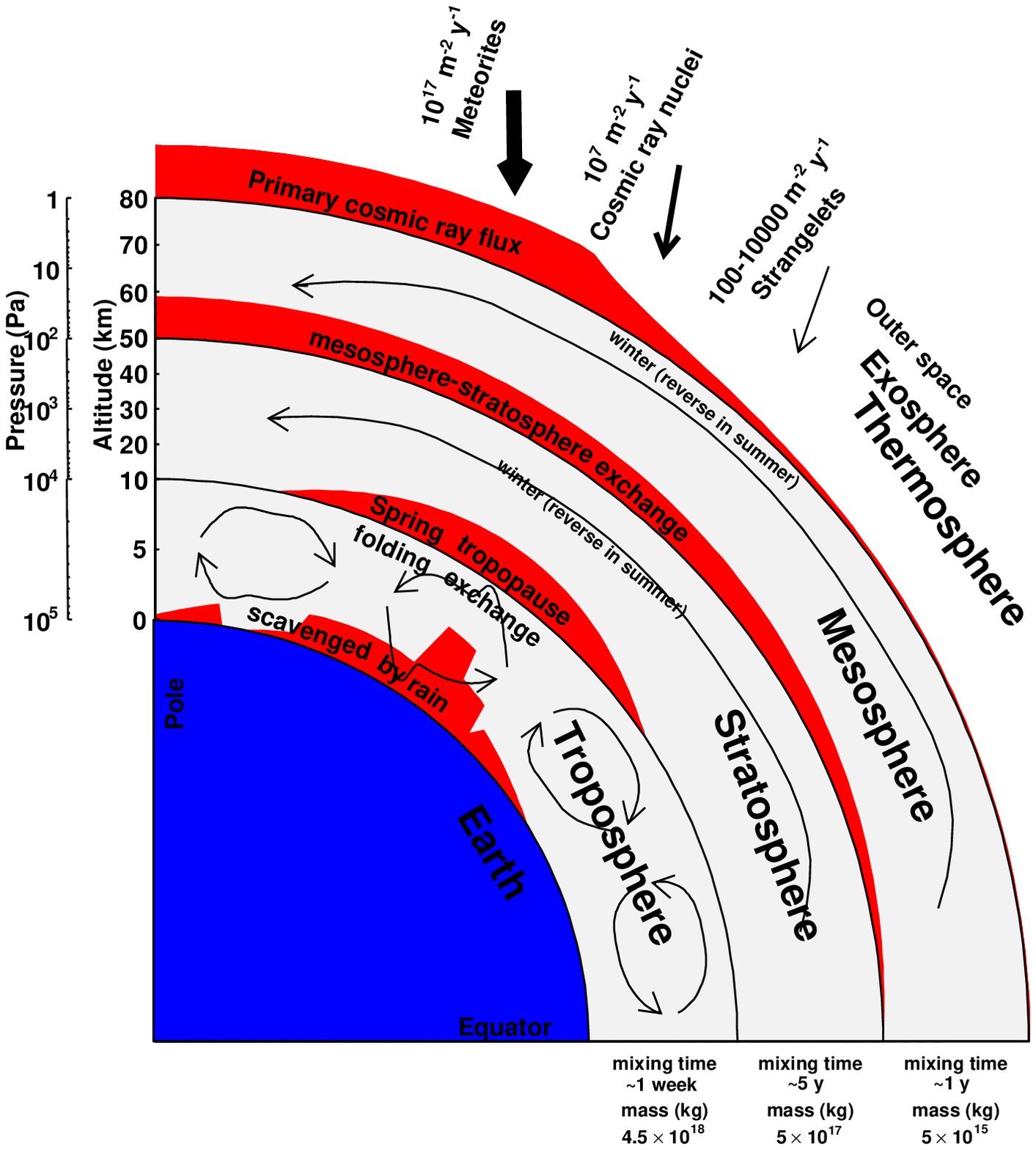}
\end{center}
\caption{Schematic cross section of the Earth's atmosphere.  The troposphere, stratosphere, and mesosphere are shown, along with their typical winter circulation patterns.  A scale on the right gives typical pressures and (mean) temperatures as a function of altitude, from \cite{ICAO}.  The grey areas show schematically where the strangelet flux crosses each boundary; note that the flux onto the Earth's surface is poorly studied, but presumed to peak around midlatitude, with regional variation which depends on precipitation.}\label{figAtm}
\end{figure*}

The mesosphere is mixed vertically by convection, with a timescale of order $\sim$1 y, as well as north-south in a ``sloshing'' motion---towards the poles in winter, towards the equator in summer.   Since most cosmic rays stop in the mesosphere and must participate in this circulation, we expect that strangelets do \emph{not} descend through the atmosphere primarily at the geomagnetic poles.  Before crossing from the mesosphere to the stratosphere, they redistribute somewhat evenly over all latitudes.  The extent of the redistribution is unclear.  

Below the mesosphere is the stratosphere ($\sim 10$--$50$ km).  The stratosphere does not convect, since the air temperature increases with altitude; metal atoms and clusters injected at the top of the stratosphere may take $\sim 5$ y to descend to the base of the stratosphere, called the ``tropopause''.  Other than meteoritic and cosmic-ray metals, the stratosphere is extremely clean; water vapor\footnote{The lower stratosphere has typically $\sim 10^{-6}$ H$_2$O.}, dust, and soot are nearly absent.  This is due to the lack of air circulation across most of the tropopause.   Ground-level air, carrying terrestrial dust, enters the stratosphere only in the tropics; tropical rainfall removes most of these contaminants as the air rises.  This air flows gradually towards the poles, in particular towards the winter pole.  Due to this pattern, we expect condensible material in the stratosphere, including metallic strangelets, cosmic ray metals, and meteorite dust, to accumulate in the polar regions towards springtime; we discuss this point further below.  In the spring, ``tropopause folds'', primarily at midlatitudes\cite{craig}, mix lower-stratospheric material back into the troposphere. 

Air circulation in the troposphere is rapid and complex.  Complete vertical turnover of the troposphere occurs on a scale of weeks; we note that this precludes any gravitational sorting of gaseous strangelets.  Water vapor contents are high ($0.01$--$10^{-4}$ by volume), and dust and soot levels are high and higly variable.  Non-gaseous materials in the troposphere are mostly scavenged by rain and snow.   

\subsection{Strange noble gases}

A strangelet with charge 2, 10, 18, 36, 54, or 86 will have the chemical properties of helium, neon, argon, krypton, xenon, or radon respectively.  These strange noble gas atoms will stay in the atmosphere forever unless a) lost into space or b) dissolved in water or rock and subducted.   Both removal effects are minor, so we expect that the atmosphere has accumulated most of the strangelets from 4.5 billion years of CR flux.  Thus, the atom abundance of a noble-gas-like strangelet in the bulk atmosphere is $c_{\mathrm{gas}} = F \times 3 \times 10^{-20}$.   This concentration factor is as high as that of the moon.

As discussed in Section~\ref{intro}, the concentration of strangelets in a pure element sample is inversely proportional to the normal element's abundance.  In particular, He and Xe are quite scarce in the atmosphere, so \str He and \str Xe nuclei may have high concentrations\footnote{In the Earth's atmosphere, helium-like strangelets are gravitationally bound while ordinary $^4$He is not.  This phenomenon means that the Earth's $^s$He/$^4$He ratio is greatly enhanced relative to the \emph{primordial} $^s$He/$^4$He ratio\cite{mueller}.   However, in terms of cosmic-ray accumulation, the Earth's $^s$He/$^4$He ratio is, as for all other noble gases, simply the integrated flux of CR $^s$He divided by the total modern amount of background $^4$He.   The fact that the modern $^4$He reservoir turns over rapidly is immaterial, as long as we recognize that CR $^s$He, once collected, does not turn over.}.  A \str Rn atom in the atmosphere would be extremely unique, because all ordinary Rn is short-lived.  The atmospheric abundances and strangelet concentrations for each noble gas are summarized in Table~\ref{TableNG}.

\begin{table}
\begin{center}
\begin{tabular}{|c| r @{$\times$}l |c|c|r  @{$\times$}l|r @{$\times$}l|}
\hline
Element & \multicolumn{2}{c|}{atmospheric} &     \multicolumn{2}{c|}{str. flux, $m^{-2}y^{-1}$}   & \multicolumn{4}{c|}{str. concentration} \\ 
        & \multicolumn{2}{c|}{abundance}   & CFL  & bag  &  \multicolumn{2}{c|}{CFL} & \multicolumn{2}{c|}{bag}\\

 \hline
He & $ 5 $ & $ 10^{-6} $& 110000& 97000& $ 2 $ & $ 10^{-10} $& $ 2 $ & $ 10^{-10} $\\
Ne & $ 1.8 $ & $ 10^{-5} $& 3900 &6600& $ 2 $ & $ 10^{-12} $ &$ 4 $ & $ 10^{-12} $\\
Ar & $ 9.3 $ & $ 10^{-3} $ &1150& 2500& $ 1 $ & $ 10^{-15} $& $ 3 $ & $ 10^{-15} $\\
Kr & $ 1 $ & $ 10^{-6} $ &270 &780 &$ 3 $ & $ 10^{-12} $ &$ 8 $ & $ 10^{-12} $\\
Xe & $ 9 $ & $ 10^{-10} $ & 120 & 400 & $ 1 $ & $ 10^{-9} $ & $ 4 $ & $ 10^{-9} $\\
Rn & \multicolumn{2}{c|}{negl.} & 45 &140& \multicolumn{2}{c|}{$>1$}  & \multicolumn{2}{c|}{$>1$}\\
\hline
N  & \multicolumn{2}{c|}{0.79} & 8200 &12000 &$ 1 $ & $ 10^{-17} $& $ 2 $ & $ 10^{-17} $\\
$^{15}$N &  \multicolumn{2}{c|}{0.0025}& 8200 &12000 &$ 4 $ & $ 10^{-15} $& $ 6 $ & $ 10^{-15} $\\
O &  \multicolumn{2}{c|}{0.20}& 6200 &9600 &$ 3 $ & $ 10^{-19} $ &$ 5 $ & $ 10^{-19} $\\
$^{18}$O &  \multicolumn{2}{c|}{0.0004}& 6200& 9600 &$ 1 $ & $ 10^{-16} $ &$ 2 $ & $ 10^{-16} $\\
\hline
\end{tabular}

\caption{Strangelets among gases in the atmosphere.  We list the atom abundance of each ``normal'' noble gas in the atmosphere.  We give the flux of CFL strangelets and bag-model strangelets according to Equation~\ref{eqFlux}, in mean number per m$^2$ per year.  Finally, we give the atom abundances of strangelets accumulated in the atmosphere assuming $4\times10^9$ y of accumulation for the noble gases, and N and O behavior as described in Section~\ref{volatiles}.  We also list $^{15}$N and $^{18}$O under the assumption that all ultraheavy isotopes are retained in commercially available heavy isotopic N and O.}\label{TableNG}
\end{center}
\end{table}

\subsection{Strange volatile elements}\label{volatiles}
Several elements, which we label ``volatiles'', can cycle back and forth between the atmosphere, ocean, and crust: H, C, N, O, S, and the halogens.  A volatile strangelet resides in the atmosphere for a time longer than the atmospheric mixing time, but shorter than the age of the Earth; volatile strangelets on the ground may be re-volatilized and reenter the atmosphere. In the case of H, C, S, we do not expect strangelets to accumulate in the atmosphere; turnover is fairly rapid and there is much dilution by ordinary matter, particularly H$_2$O, CO$_2$, and SO$_X$.  The halogens case for halogens is quite similar; all halogens are present in the atmosphere, both naturally and as pollutants: fluorine as manmade fluorocarbons; sea-spray HCl in the troposphere, natural CH$_3$Cl and manmade CFCs in the stratosphere; bromine as BrO over oceans, and CH$_3$Br, CHB$_3$, and halons throughout; iodine as various iodocarbons.  The fully-halogenated species have atmospheric lifetimes of order 100 y; all other species have short lifetimes, usually $<$ 1 y, so we do not expect halogen-like strangelets to be at all abundant in atmospheric samples.  

Nitrogen is more interesting.  Most of the Earth's entire nitrogen inventory ($4 \times 10^{18}$ kg) is in the atmosphere as N$_2$\cite{jaffe92}.  Although nitrogen cycles back and forth into solid forms, mainly due to biological fixation, at a rate of $10^{11}$ kg/y, it is very slow ($10^{10}$ kg/y) to be sequestered in the crust\footnote{Nitrogen sequestration is mainly via deposition (diatom shells, etc.) on the ocean floor.  Interestingly, industrial-scale agriculture has nearly doubled the global fixation budget over the past few decades\cite{vitousek}.}. (Much of this, too, eventually reenters the atmosphere via uplift and erosion.)  Although it is difficult to project the nitrogen cycle into the distant geological past, the modern nitrogen cycle suggests that the lifetime of a \str N atom in the atmosphere is of order $5\times10^8$ years.  The effective lifetime may be much longer, depending on how much sequestered nitrate is eventually re-exposed rather than subducted.  Conservatively, taking the net \str N flux of the past $5\times10^8$ y to remain in the atmosphere, we expect a \str N concentration of $5\times10^{-18}$. 

Oxygen is one of the most abundant elements on Earth, making up much of the mass of the biosphere, oceans, crust, and mantle in the form of water, metal oxides, and organic molecules.   However, the most likely chemical fate of an an arriving \str O atom is to be incorporated into O$_2$.  O$_2$ has an atmospheric residence time of $3 \times 10^6$ y\cite{keeling93}, suggesting a \str O abundance of $10^{-19}$ in atmospheric O$_2$\footnote{Aerobic respiration turns atmospheric O$_2$ into H$_2$O---the oxygen in other biological compounds, like glucose, is derived from H$_2$O and CO$_2$---so high \str O concentrations are not expected in the biosphere\cite{biochemistry}}.  

\subsection{Metallic strangelets in the stratosphere}\label{secMetal}

Most possible strangelet charges correspond to metallic elements.  A metallic strangelet stopped in the mesosphere will mix immediately with the metals left behind by vaporizing micrometeors.  These metals agglomerate into nanometer-scale ``smoke'' clusters within a few months of arrival\cite{hunter80}.  This smoke has been observed to drift down into the lower stratosphere and collect in sulfuric acid aerosol droplets\cite{cziczo}; 1/2 of such droplets found at midlatitudes carry 0.5--1\% by mass of dissolved metal, with approximately chondritic composition.  From this, and the known flux of sulfur to the stratosphere (\mbox{100--160 $\times 10^6$ kg/y}), the flux of vaporized meteoritic material to the stratosphere is inferred to be 4--19$\times 10^6$ kg/y.\footnote{Note that this measurement specifically probes the amount of meteoritic material which is left in the stratosphere, not the total amount which reaches the Earth.}

For further discussion, let us consider $10^7$ kg/y (roughly the middle of the plausible range) to be the meteorite flux.  Of this amount, approximately 44\% is oxygen, and the remainder amounts to $1.7 \times 10^8$ moles of metals entering the stratosphere per year\footnote{Three other sources of metal are available to the stratosphere.  Major volcanic eruptions inject ash into the lower stratosphere, where it remains for $\sim 1$ year.  Recent eruptions of this power were Mt. St. Helens (1981), El Chichon (1982), Nevado del Ruiz (1985), Mt. Augustine (1986), and Mt. Pintubo (1991).  Large aboveground nuclear tests may also affect the stratosphere, but we hope these are no longer an issue.  High-altitude aircraft, including the Concorde, and rockets also contribute.}.  Thus, the atom abundance of strangelets mixed in with these metals is $c_{\mathrm{strat}}  = F \times 1.5 \times 10^{-17}$ where F is the flux from equation~\ref{eqFlux}.   This is a factor of 1000 higher than the concentrations on the moon, and $10^6$ higher than concentrations in the Earth's crust.  Due to this high concentration, stratospheric aerosols may be a target for a strangelet search, discussed below.  

We also note that the stratospheric concentration factor provides an upper limit, albeit a weak one, for cosmic-ray strangelet concentrations in rainfall or in geological samples;  no Earth environment is exposed to metallic strangelets without also being exposed to micrometeorite smoke.  

\subsection{Thermodynamic and chemical effects of strangelet mass}

We might be concerned that the unusually large isotope shifts\cite{bigeleisen_book} affecting strangelets could change their chemistry or thermodynamic behavior in such a way as to invalidate our atmospheric residence times, or else to defeat some sample collection techniques.  These effects can be estimated, albeit with considerable uncertainty.  From such a calculation, we conclude that the errors in our analysis due to isotope effects are less than a factor of 2, and thus much smaller than the other (e.g., astrophysical) uncertainties relevant to a strangelet search.

The effect of isotopic mass on vapor pressure is discussed in section~\ref{secDist}; the vapor pressure shift is very similar in principle to shifts in chemical equilibria, solubility, etc..  We show in Table~\ref{TablePhysProp} that strange gases should have only slightly depressed vapor pressures.  Although these numbers are approximate, we may conclude that strangelets are not removed from noble gases by commercial distillation plants\footnote{Strangelets may vary in concentration \emph{within} a particular noble gas fraction in a column; however, the normal operation of a commercial distillation plant---in which the ``waste'' gas from a given stage is not discarded, but rather re-injected at an earlier stage---should not maintain this separation.}.   We can also conclude that strangelets do not behave significantly differently than ordinary gases in nature.  Ordinary isotope effects are sometimes observed to cause 1\%-2\% variations in natural isotope abundances; since strangelet vapor pressure shifts are only a factor of O(10) larger, and natural processes do not typically cascade or amplify these effects, the variation in strangelet abundances between natural samples should be of order 10--50\% at most.     

\section{Aspects of strangelet searches}

Now that we have identified these environments where strangelets may be present at high concentrations, we can design experiments to search for them.  The three steps in any search are: sample collection, purification/preconcentration, and mass spectroscopy.   

We would like to emphasize that we do \emph{not} know, if strangelets are stable: what is the minimum mass $A_t$? What is the exact charge/mass relation?  What is the distribution of masses produced by strange star collisions?  What is the mass distribution for a sample with a given Z?  In order to comprehensively address the question of the existence of strangelets, we should perform experiments over a wide range of charges, and each search should be sensitive to a wide range of possible masses.  Although some particular searches stand out for their high sensitivity, we wish to investigate more of the periodic table in order to cover the search space completely.  

\subsection{Past atmospheric strangelet searches}\label{past}

Several searches for strangelets in meteorite\cite{hemmick}, Earth rock\cite{brugger}\cite{perillo} and in cosmic rays \cite{lowder91} have set upper abundance limits.  Stellar structure places strong constraints on the strangelet abundance in the Sun\cite{takahashi}, in particular probing \str H and \str He.  The current state of experiments is reviewed by Klingenberg\cite{Klingenberg}.  We describe the results most relevant to atmospheric strangelets.    

Mueller et.\ al.\ \cite{mueller} searched for \str He in the Earth's atmosphere.   This search, using absorbtion spectroscopy, limits the isotopic abundance of \str He to be $< 10^{-8}$ for strangelets of mass $> 20$ amu.  This is not far off from the predictions of the Madsen model. Vandegriff et.\ al.\cite{vandegriff} searched for \str He in the mass range 42--82 and found abundances $< 2 \times 10^5$.   Both authors focus on primordial strangelets, for which the \str He/He ratio on Earth is enhanced by a factor of $\sim 10^7$ over the average galactic \str He/He ratio by gravitational trapping; this enhancement factor is not relevant for cosmic-ray strangelets.  Interpreted as a limit on strangelets in CR, Mueller result constrains the strange star production rate of \str He to be $< 5\times10^{-9} \sun$ y$^{-1}$.

Hemmick et.\ al.\ \cite{hemmick} searched for \str O using a commercial $^{18}$O sample, as discussed in Section~\ref{samples}.  They find strangelet abundances no greater than $4 \times 10^{-17}$ in bulk oxygen, or $1 \times 10^{-15}$ in their enriched sample\footnote{We presume that their enriched sample of $^{18}$O, like ours, comes from distillation of CO containing atmospheric O.} using a tandem accelerator with an all-electrostatic beam line and a segmented gas chamber.  This constrains the strange star production rate of \str O to be $< 3\times10^{-7} {\sun}$ y$^{-1}$.

Holt et.\ al.\ \cite{holt} searched for ``collapsed'' Rn nuclei using an atmospheric Xe sample.  They collected Rn from the equivalent of $10^4$ l of atmospheric Xe, using the reaction (IF$_6)^+$(SbF$_6)^-$~+~Rn$_{(g)}$~$\rightarrow$~(RnF)$^+$(SbF$_6$)$^-_{(s)}$~+~IF$_5$\cite{stein1982}.  This sample was illuminated with a high flux of thermal neutrons, and monitored for the emission of 30--250 MeV gamma rays.   They determined the number of ``collapsed'' Rn atoms to be $< 3 \times 10^{10}$.  If this result applied to \str Rn, it would limit the strange star production rate of \str Rn to be $< 2\times 10^{-17} {\sun}$ y$^{-1}$.  However, we do \emph{not} expect such high-energy gamma emission from strangelet neutron capture\cite{jaffe_radioactivity}, so this limit does not apply.  Nevertheless, this search illustrates the potential for a \str Rn search to limit strangelet production rates.   

\subsection{Obtaining samples}\label{samples}

The standard technique for obtaining noble gases is fractional distillation of air.  It appears that most strange gases will distill out along with the normal gases: \str Ne with Ne, \str Kr with Kr, etc.\footnote{It should be noted that the details of commercial distillation plants are proprietary; one can imagine design details, albeit odd ones, that would cause strangelets to be discarded.}.  \str Rn comes out in the Xe fraction.  To search for strange noble gases, it should be sufficient start with any commercial gas sample.

The exception is helium; commercial helium supplies are extracted from natural gas, where it accumulates largely due to alpha decay rather than due to any primordial or atmospheric source.  Therefore, there is no good reason for these samples to contain cosmic-ray strangelets.  A strangelet search must be performed on a specially-collected atmospheric He sample\cite{mueller}.  

 We also note that $^{15}$N (natural abundance 0.0037) is extracted from atmospheric N$_2$, by distillation or chemical exchange, on a commercial scale\cite{sigmaaldrich}.  Strangelets will follow the heavy fraction in this distillation,  so the \str N abundance will be 270 times greater in a  $^{15}$N sample than in a natural N sample.   99\% pure $^{18}$O (natural abundance 0.002) is separated commercially via distillation of CO\cite{sigmaaldrich}; this concentrate will have 500 times the strangelet concentration of natural O.   We include O, N and concentrated $^{18}$O and $^{15}$N in Table~\ref{TableNG}.

In addition, \str N should be present at atmospheric concentrations in most solid nitrogen samples, which come from the modern atmosphere via the Haber process or via the biosphere.  (The same is not true of oxygen.)

\subsubsection{Stratospheric metals}

The idea of searching for strangelets in stratospheric metals is somewhat speculative; the primary difficulty is obtaining a large enough material sample.  The sample should be gathered from the low polar stratosphere (10-20 km altitude) in late winter.   An appropriate collecting device could be launched on a weather balloon or a high-altitude research aircraft.    

A large, lightweight electrostatic precipitator could in principle collect aerosol droplets in large quantities from stratospheric air.   Although power, weight, and efficiencies need to be evaluated, if one could construct a precipitator which collected all of the aerosol materials in a 1 m$^2$ area, while making a single pass through a 20 km aerosol layer in the stratosphere, would collect about 1 mg of aerosols\cite{arnold98}, carrying 10$^{15}$--10$^{16}$ atoms of metals\cite{cziczo} and of order $\sim$ 10--100 strangelets.  Clearly many such flights, or a few more complex missions, would be needed to collect an analyzable sample.  However, we emphasize that the stranglet concentration in this sample, if not the sample size, is within the reach of some modern mass spectrometers.  Furthermore, we note that such a sample may contain any of a large number of strangelet charges and masses, in constrast to gas-based searches which require some strangelets to have charge 2, 7, 8, 10, 18, 36, etc.  

\subsubsection{Distillation}\label{secDist}

Distillation separates isotopes by taking advantage of the small differences in vapor pressure between isotopic species.  This difference can be calculated, for monatomic gases, from the element's Debye temperature\cite{van_hook_book}.  For a gas with a light isotope of mass $M'$, vapor pressure $P'$, and Debye temperature $T_d$, and a heavy isotope with $M$ and $P$, the vapor pressure at temperature $T$ obeys

\begin{equation}\label{eqPVap}
\mathrm{ln}(\frac{P'}{P}) = (\frac{T_d}{T})^2 \frac{M'}{24}(\frac{1}{M'}-\frac{1}{M})
\end{equation}

\begin{table}
\begin{center}
\begin{tabular}{|c|c|c|c|c|c|}
\hline
Element & T$_d$ & \multicolumn{3}{|c|}{Ordinary isotope shifts} & Strangelet\\
        &  K    &       & measured & theory                &   prediction  \\
        &       &        & P$_l$/P$_h$ & P$_l$/P$_h$         &   P$_l$/P$_{str}$  \\

\hline
Ne &  75 &  $^{20}$Ne/$^{22}$Ne  & 1.044\cite{keesom}  & 1.03 &   1.38 \\ 
Ar &  92 &  $^{36}$Ar/$^{40}$Ar  & 1.006\cite{clusius}  & 1.004 &   1.05 \\
Kr &  72 &  $^{80}$Kr/$^{84}$Kr  & 1.0007\cite{kr_vpie}     & 1.0007 &   1.01 \\ 
Xe &  64 & $^{130}$Xe/$^{136}$Xe & 1.0001\cite{jancso}      & 1.0002 &   1.006 \\
Rn &  75 &              ---        &   ---     & --- & 1.005 \\   
\hline
O$_2$ & 93 & $^{16}$O$^{16}$O/$^{18}$O$^{16}$O  &  1.01\cite{jancso}      & 1.003  &   1.06 \\
N$_2$ & 83 & $^{14}$N$^{14}$N/$^{15}$N$^{14}$N  & 1.005\cite{jancso}       & 1.002  &   1.07 \\
CO & $\sim$81 & $^{12}$C$^{16}$O/$^{13}$C$^{16}$O  & 1.01\cite{london1958} & 1.02 & 1.05 \\
\hline
\end{tabular}
\caption{Isotope effects for strange gases.   The second column gives the Debye temperature T$_d$ (K) in K.   The fourth and fifth columns show the vapor-pressure shift for an ordinary isotope system, with both the experimental values and the Eq.~\ref{eqPVap} prediction.  In the sixth colum, we give a prediction for the vapor pressure change P/P$_{str}$ for an arbitrarily massive strangelet, calculated at the boiling point using Eq.~\ref{eqPVap}.  Eq.~\ref{eqPVap} is also used for N$_2$, O$_2$, and CO, but the results should be trusted only at the order-of-magnitude level.}\label{TablePhysProp}
\end{center}
\end{table}

Compared to monatomic gases, the calculation of vapor-pressure isotope effects in polyatomic molecules is difficult; vibrational and rotational excitations make large contributions to the isotope shift.  For example, the pressure change at 77K, relative to $^{12}$C$^{16}$O, differs by a factor of three between $^{12}$C$^{18}$O, $^{13}$C$^{17}$O, and $^{14}$C$^{16}$O, all of which have the same mass\cite{van_hook_book}.  We make no attempt to calculate them, and use Eq.~\ref{eqPVap} to give a ballpark figure.  

For ideal mixtures, distillation relies on the elementary separation factor $q_0 = p_1/p_2$, where $p_x$ is the vapor pressure of isotope $x$ in the pure state.  In general, the ultimate separation power of a column is a complicated function of this separation factor.   In the special case of strangelets, where the heavy species always has a very small abundance, the behavior simplifies.   In a distillation column (or any repeated batch process) with $N$ equilibration steps, the maximum obtainable concentration factor is $c_{max} = q_0^N$.   

For the purposes of this study, rather than designing a distillation column from first principles and optimizing it for strangelet separation, we will consider several columns used for isotope separation in the past.   Consider the column used by Johns and London\cite{london1958} for the production of $^{13}$C from CO.   A column 32 feet long used 600 equilibration stages to enrich $^{13}$C from a natural abundance of 0.011\% to approximately 0.6\%, a factor of 400.   The feed rate was 625g/day, and the enriched product was extracted at 0.4g/day.   The single-stage separation factor is 1.01.   Running the same column to separate \str Xe from Xe (single-stage separation factor 1.006 at the boiling point) would give an enrichment of a factor of 40.   Running on \str Kr and Kr (single-stage separation factor 1.01) would give an enrichment of $10^2$--$10^3$\footnote{A small uncertainty on the single-stage separation leads to a very large uncertaintly in the column performance.}   For lighter gases, the single-stage enrichments are larger; operating the column at the original feed rate would give a maximum enrichment factor of $1.5\times10^3$ (a number which depends on the amount of material in the reboiler to which the strangelets are transported.) Reducing the enriched product withdrawal rate, or increasing the feed rate, would increase the enrichment factor approximately linearly.  

A smaller column, like that used by Clusius and Mayer \cite{clusius} for the separation of $^{36}$Ar, would be entirely adequate for work with \str Ne, \str Ar, \str N, or \str O.  This column produced an enrichment factor of 2 using 160 stages with a single-stage separation of 1.005; running this column on \str Ar would give a strangelet enrichment of $10^3$.  The column produced 0.7 moles of enriched product per day.   

For \str Ne, due to the very low temperatures involved, an even smaller column would be desired; an apparatus with 20 stages would be sufficient to concentrate strangelets by a factor ~1000.   Even the fairly small apparatus of Keesom et. al. \cite{keesom}, with 60 separation stages, is overdesigned for \str Ne separation; as operated on 22Ne, this column would enrich \str Ne by a factor $\sim500$ in a single pass (~4d), and would increase that enrichment factor approximately linearly with additional running time.

\begin{table}
\begin{center}
\begin{tabular}{|c|c|c|c|c|c|}
\hline
Element & P$_l$/P$_{str}$ & atm. conc. & distillation & concentration & strangelet\\  
        & at T$_b$ &        & design    & in reboiler  & rate (mol/day) \\
\hline
Ne &  1.37 & 2e-12 & Keesom column & 1.6e-10/d & 1e-11  \\
Ar &  1.05 & 1e-15 & Clusius column &  9E-15/d & 6E-15\\ 
Kr &  1.01 & 3e-12 & London column & 5e-9/d  & 6e-11\\
Xe &  1.006 & 1e-9 & London column & 4e-8 (eq)  & 5e-10\\
O$_2$ & 1.06 & 5e-19 & Clusius column & 5e-18/d  & 3e-18\\
N$_2$ & 1.07 & 2e-17 & Clusius column & 2e-16/d  & 1.2e-16\\
\hline
\end{tabular}
\caption{Expected behavior of strangelets in some documented distillations columns; Keesom et. al.\cite{keesom}, Clusius and Mayer,\cite{clusius}, and London\cite{london1958}.   We extrapolate the performance of each column, using its response to ordinary isotope shifts, to its performance on strangelets as calculated in~\ref{TablePhysProp}.  For columns with large separation factors, the concentration at the bottom of the column should increase with time; for the smaller separation factor in Xe, the column reaches equilibrium after concentrating \str Xe by a factor of 40. }\label{TableColumns}
\end{center}
\end{table}

\subsubsection{Thermal diffusion separation}
Thermal diffusion separation\cite{vasaru} is very powerful and efficient at separating heavy gases from light mixtures.  A typical column\cite{london}\cite{spindel_book} might consist of a hollow tube 25mm in diameter and 3m long, water-cooled on the outside, with a coaxial wire electrically heated to 600K.  Light isotopes preferentially migrate towards the hotter gas and are swept upwards by convection; heavy isotopes concentrate at the bottom.   It is straightforward to operate several short columns in series so that they behave as a single long column. 

Consider a column of length $L$, filled with a gas of molecular mass $m$, and a trace concentration $c_0$ of strangelets of mass $M$.  At equilibrium, the strangelet concentration $c_L$ at the bottom of the column is increased by 

\begin{equation}
q_e \equiv \frac{c_L}{c_0} = k e^{L/\lambda}
\end{equation}

where the characteristic length $\lambda$ scales approximately as $\lambda  \propto (M-m)/(M+m)$, and $k$ depends in detail on diffusion parameters, temperature, viscosity, etc.\cite{london}   For the heavier gases, this equation predicts extremely large values for $q_e$ even for fairly short columns.  In practice, very large separation constants suggest that strangelets are transported down the column unidirectionally.  There is unavoidably some dead space at the bottom of the column; the column's whole inventory of strangelets will accumulate in the dead space in the equilibration time $\tau$.  This allows us to compute the strangelet production rate $P$, the number of strangelets transported down the column per day.   

The final strangelet concentration at the bottom of the column will depend on $q_e$, $P$, the run time, and the amount of dead space from which the final sample is drawn.   The dead space consideration may be removed by mixing the initial sample with an intermediate-mass buffer gas.  For example, consider a long column containing \str O$^{16}$O and $^{18}$O$_2$.   A sample drawn from the bottom of the column may contain 60 times as much \str O$^{16}$O as the source gas, but the bulk of the sample will still be $^{18}$O$_2$.   Compare this to a column containing \str O$^{16}$O , $^{18}$O$_2$ and Kr as a buffer gas.  In this case, a bottom sample will contain \str O$^{16}$O mixed with bulk Kr, to the near-complete exclusion of $^{18}$O$_2$.   This sample is subject to additional chemical purification.  This allows the preparation of extremely concentrated strangelet samples from almost any gas, with appropriate choice of buffer.

\begin{table}
\begin{tabular}{|r|r @{$\times$} l |c|c|c|c|c|}
\hline 
Gas & \multicolumn{2}{c}{strangelet} & $\lambda$ & $q_e$ & $\tau$ & $P$ \\     
    & \multicolumn{2}{c}{$c_0$ (CFL)} &  (cm)    & (3m) &   (d)    & (d$^{-1}$)\\
\hline
He & $2$ & $ 10^{-10}$ & 550 & 2 & 9 &  $10^{12}$\\
Ne & $2$ & $ 10^{-12}$ & 90 &  24 & 0.35 & $10^{11}$\\ 
$^{15}$N$^2$ & $4$ & $ 10^{-15}$ & 70  & 60 & 0.5 & $10^{9}$ \\
$^{18}$O$^2$ & $1.5$ & $ 10^{-16}$ & 70 & 60 & 0.5 & $10^{7}$\\
Ar & $1$ & $ 10^{-15}$ & 50 & 460 & 0.14 & $10^8$\\
Kr & $3$ & $ 10^{-12}$ & 30 & 7000 & 0.1 & $10^{12}$\\
Xe & $1$ & $ 10^{-9}$ & 15 & $5\times10^8$ & 0.08 & $10^{14}$\\
\hline
\end{tabular}
\caption{Behavior of strangelets in a thermal diffusion column.  These parameters are calculated using the planar approximation of \cite{spindel_book} for a hot-wire column of radius 1.2cm, wall temperature 300K, wire temperature 600K, and length 3m.  $c_0$ is the strangelet atom abundance for this species (for CFL strangelets only).  $\lambda$ is the characteristic length scale for this gas and this column.  $q_e$ is the theoretical concentration enhancement, at equilibrium, at the bottom of a column.  $\tau$ is column equilibration time in days.  $P$ is the approximate number of strangelets transported down the column per day, assuming atmospheric abundances as Equation~\ref{eqFlux}.}\label{tableThermal}
\end{table}

\subsubsection{Chemical separation of Rn}
We can collect very large atmospheric Rn samples, including \str Rn, by extraction from Xe over (IF$_6$)$^+$(SbF$_6$)$^-$, as described in section~\ref{past}.  An even simpler technique would be chromatography on activated carbon. A stream of mixed Xe and Rn is passed over activated carbon at room temperature.  Rn is strongly retained\cite{cohen}.  Later, the Rn is released by heating under a stream of He or N$_2$.  In either case, 1 mole of Xe (22.4 lSTP) with the Table~\ref{TableNG} strangelet concentrations would yield $3\times10^{14}$ \str Rn atoms.

\subsubsection{Isotope separation by chromatography}\label{chrom}

If it is necessary to pre-concentrate strangelets from a stratospheric metal sample, this must be done with very high efficiency for the strangelets; very small initial samples cannot be wasted.  However, since the strangelet concentrations in these samples are high to begin with, we may be content with a low-throughput technique like chromatography.  Ion-exchange chromatography is known to show small isotope separation effects.   We show some data in Table~\ref{tabLCCE}; more can be found in \cite{london} and \cite{fujii}.  Assuming these effects to be due solely to nuclear mass effects\footnote{As always, this assumption may be distorted by nuclear size, spin, and quantum effects\cite{ishida}.}, and following the Bigeleisen-Mayer prescription\cite{bigeleisen}\cite{bigeleisen_book}, we extrapolate these small effects to strangelets by assuming that the isotope effect $\epsilon$ (basically, the fractional difference in the solubility product between isotopes of mass $M$ and $m$) depends on the masses as $\epsilon = \epsilon_0 \frac{M-m}{M\times m}$.  The separation power of a chromatography column scales of length $L$ scales as $L \epsilon^2$.  Large separation factors, like those predicted for \str Li and \str Mg, would allow near-complete strangelet purification in a single column.  Smaller separation factors, as for \str V--\str U, permit concentration by factors of 10-100 in simple setup.  In principle, such a process can purify and recover nearly 100\% of a strangelet sample, although this requires testing.  

\begin{table}
\begin{tabular}{|r r|r|r|}

\hline
Element & [ref] & $\epsilon_{obs}$ & $\epsilon_{\mathrm{str}}$ \\
\hline
$^6$Li/$^7$Li & \cite{li_chrom} & .047 & 0.22\\
$^{24}$Mg/$^{26}$Mg & \cite{mg_chrom} & .014 & 0.15 \\
$^{50}$V/$^{51}$V & \cite{zhang_v} & $2\times10^{-4}$ & 0.01\\
$^{64}$Zn/$^{66}$Zn & \cite{zn_chrom} & .001 & 0.03\\
$^{134}$Ba/$^{138}$Ba & \cite{ba_chrom} & $3\times10^{-4}$ & 0.01 \\
$^{235}$U/$^{238}$U & \cite{u_chrom} & $3\times10^{-4}$ & 0.02\\
\hline
\end{tabular}

\caption{Sample experimental data on isotope separation by chromatography.  For each isotope, the measured isotope partition factor $\epsilon_{obs}$ is extrapolated to $\epsilon_\mathrm{str}$ at the CFL strangelet mass.}\label{tabLCCE}
\end{table}

\paragraph{Electromagnetic separation}
It is always possible to separate gas-phase ions by mass using electromagnetic fields.  Traditional ``calutrons'' separate isotopes using a spark ion source, an accelerating potential of order 10--20 kV, and large-area ($>$1m$^2$) magnetic field.  Ions with different m/z have different trajectories in the magnetic field; beams of various masses are made to implant in collector foils at different positions, from which the atoms are later extracted.  Consider a calutron which produces a beam of 10 $\mu$A of ions from a strangelet-containing oxygen sample.  Supplying this source with straight commercial $^{18}$O, with the nominal strangelet concentration, produces $10^5$ separated strangelets per month.   The collection system is necessarily complicated, in a search experiment, by the fact that the strangelet mass is unknown.   The purity of the resulting sample will be limited by, among other things, the nonzero beam size and halo, which will send some fraction of the non-strange ions into the strangelet collector area.   This suggests that electromagnetic separation is not promising compared to thermal diffusion separation.   

Because typical ion sources have very low source ionization/transmission efficiencies ($\sim10^{-5}$), electromagnetic separation is unsuitable for the already-small stratospheric metal samples.  It should be kept in mind for future terrestrial and lunar work.

\section{Mass spectroscopy}
After collecting and pre-concentrating a strangelet-bearing sample, one must perform mass spectroscopy to try to identify strangelets.   A detailed survey of the spectroscopy techniques available is beyond the scope of this paper.  However, we observe that many extremely-sensitive techniques are available off the shelf.   There are three major factors influencing one's choice of mass spectrometer: sensitivity, sample consumption, and mass range.  For example, accelerator mass spectrometry (AMS) is often capable of detecting elements or isotopes at concentrations of 10$^{-14}$, but consumes comparatively large amounts of sample, and is only sensitive to $\sim$ 1 amu at a time.   Fourier transform ion cyclotron resonance mass spectroscopy (FTICR-MS) can process very tiny samples, and observes a wide range of candidate masses, but cannot deal with isotopic abundances below about $10^{-6}$.   Table~\ref{tabMS} gives a brief overview of some mass spectroscopy techniques which could be used for strangelet searches.  

\begin{table}
\begin{center}
\begin{tabular}{|m{5cm}|m{1.5cm}|m{1.5cm}|m{1.5cm}|m{5cm}|}
\hline
Technique                                      &  Sens.  &  Eff. &  Range & Comments\\
\hline
\hline
Accelerator mass spectrometry (AMS) \cite{hemmick}\cite{sandweiss}                   & $10^{-17}$    & $< 10^{-5}$                     & $< 1$ amu & Ultra-high sensitivity possible with extra background rejection at detector\\ 
\hline
Atom Trap Trace  Analysis (ATTA) \cite{atta_nimb} & $>10^{-12}$   & $10^{-8}$                      & Large & Well-developed for Kr \\ 
\hline
Inductively coupled  plasma (ICP) \cite{jarvis92}                           & $10^{-14}$    & $< 10^{-5}$           & $<1$amu & Quoted efficiency for O, N; worse for noble gas, much higher for metals\\
\hline
Fourier-transform ion  cyclotron resonance (FTICR) \cite{marshall85}\cite{belov}\cite{tisato04}      & $10^{-9}$    & $>10^{-3}$  & Large  & Good for tiny but concentrated samples; organic molecules only\\
\hline
Optical isotope shift  spectroscopy \cite{mueller} & $10^{-8}$  &  ---                             & Large  & Easy detection of Rn?\\ 
\hline
Resonant ionization mass  spectroscopy \cite{wendt00}\cite{sarisa_book} & $10^{-13}$  &  high  &  Large  & Best efficiencies for metals only\\ 
\hline
\end{tabular}
\end{center}
\caption{Overview of some mass-spectroscopy techniques, including their sensitivity (minimum detectable concentrations), efficiency (fraction of sample atoms actually sensed), and the range of masses accesible in one pass.   Techniques like AMS and ICP would be best suited to light strangelets, for which the mass scan range is not too large.  ATTA is somewhat element-specific; adapting it to non-noble gases would likely reduce the efficiency.   FTICR-MS has been used for extremely tiny samples, and has the benefit of ultra-high mass resolution, but its dynamic range is small, and some chemical manipulation is required to feed strangelets into the preferred nano-electrospray sources.  Optical spectroscopy of isotope shifts is a promising way to spot rare components in a heterogenous sample (like the samples one gets from a buffered thermal diffusion column), but is less sensitive for finding a rare isotope in a large sample of the same element. Resonant ionization mass spectroscopy has all of the desired properties---high sensitivity, high efficiency, wide mass range---but only for metals with appropriate ionization schemes; for noble gases, N, and O, the sensitivity is much lower.}      \label{tabMS}
\end{table}

\section{Summary and conclusions}

The Earth's atmosphere serves as a beam dump for cosmic ray strangelets.  In the upper atmosphere, recently-fallen strangelets accumulate and concentrate in aerosol droplets, free from contamination by normal nuclei.  In the atmosphere as a whole, strange isotopes of noble gases, N, and O should have accumulated to quite high concentrations.  These concentrations are within the reach of modern mass spectroscopy, even if strangelet fluxes are quite small.  

We sum up strangelet accumulation in a given environment by a ``concentration factor'', shown in Table~\ref{tabConc}.  This is the per-atom abundance of strangelets in this environment, divided by the flux in m$^{-2}$y$^{-1}$.  This factor shows that stratospheric aerosols yield the highest strangelet concentrations known; the Earth's atmosphere and the moon are equally good places to obtain large quantities of less-concentrated strangelets.  The Earth's crust has a very low concentration factor, so effective strangelet searches in rock will require additional effort. 

\begin{table}
\begin{tabular}{|l|c|c|} 
\hline
Sample  & $c/F$ & Search\\
    & (m$^2$ y) & candidates \\
\hline  
Earth crust  & $1\times10^{-23}$ & rare metals (\str Tc, \str Pu) \\
Lunar soil   & $3\times10^{-20}$ & any metal, \str O\\
Meteorite   & $\sim 1\times10^{-22}$ & any metal, \str O\\
Noble gases  & $3\times10^{-20}$ & \str He, \str N, \str Xe, \str Rn\\
Stratospheric metals & $1.5\times10^{-17}$ & any metal\\
\hline
\end{tabular}
\caption{Summary of strangelet accumulations in various environments.  The figure of merit is the ``concentration factor'' $c/F$, per-atom concentration divided by flux per m$^2$ y.  In the third column, we mention some particularly interesting or powerful search channels in each environment.} \label{tabConc} 
\end{table}

\begin{figure}
\begin{center}
\epsfxsize=6.0in
\epsfbox{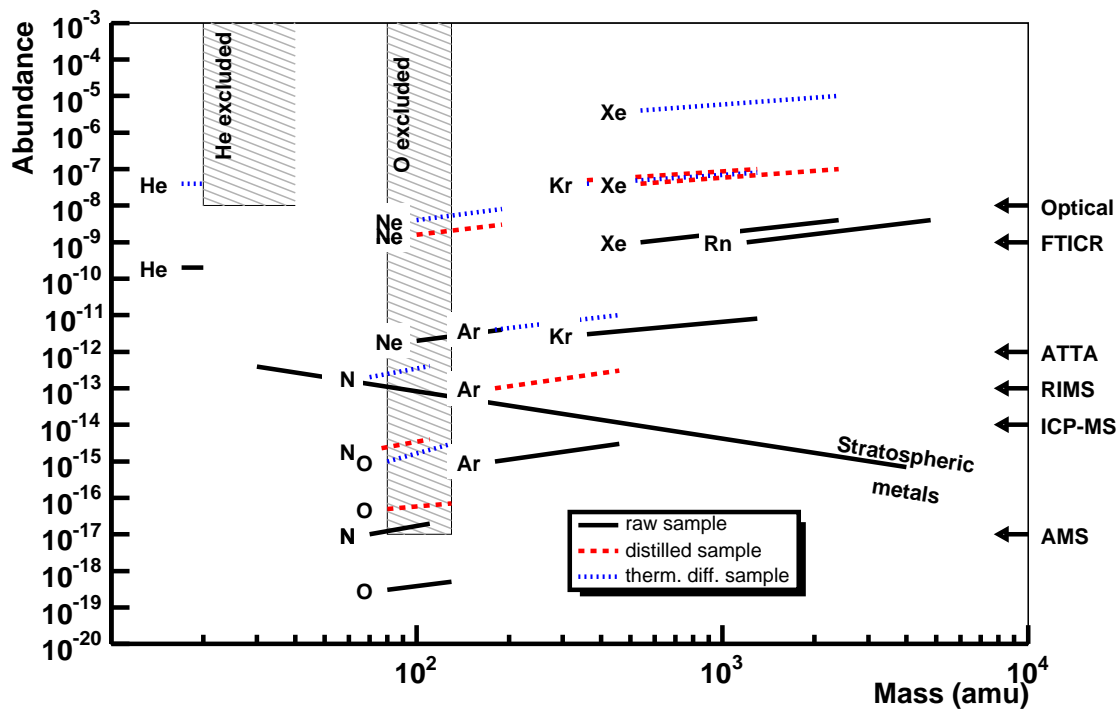}
\end{center}
\caption{Summary of expected strangelet concentrations in the atmosphere.  Black lines show the expected strangelet abundances in pure atmospheric samples (with the exception of Rn, which is given as an abundance in a Xe sample).  Additionally, we show the concentration obtainable by simple isotope separation processes on pure atmospheric gases.  The dashed lines show the expected abundance of strangelets in a 10-day run of the distillation columns in Table~\ref{TableColumns}.  The dotted lines show the expected abundance of strangelets in a 100-day run of the thermal diffusion column in Table~\ref{tableThermal}, assuming a 100-cc final sample and no medium-weight buffer gas.   The hatched regions are excluded by past atmospheric experiments.   The arrows on the right are suggestions for the limiting sensitivity of some mass-spectroscopy techniques.}
\label{figExpSummaryPlot}
\end{figure}

Taken together, these suggestions constitute a fairly comprehensive strangelet search program which could be executed mainly with off-the-shelf equipment.  An optical search for \str Rn offers an excellent opportunity to discover high-mass strangelets.  A deep search for low-mass strangelets can be carried out on atmospheric \str He, \str N, \str O, and \str Ne, by carrying out thermal-diffusion separation and accelerator mass spectroscopy.   Intermediate-mass strangelets might be found in strange transition metals in the stratosphere, searched for by resonance ionization mass spectroscopy, or in \str Kr, with an ATTA search.   An overview of the expected concentrations is shown in Figure~\ref{figExpSummaryPlot}.   This illustrates that a very large strangelet search space can be addressed using the atmospheric sampling and preconcentration techniques described here.   

The author would like to thank Sarah Bagby, Edward Boyle, Alan Davison, Peter Fisher, Ben Lane, David Mohrig, Alan Plumb, David Pritchard, Joseph L.\ Smith, and Forest White of MIT, Dick Majka and Jack Sandweiss of Yale, Jes Madsen of Aarhus, Alan Marshall of Florida State University, and Wallis Calaway and Zheng-Tian Lu of Argonne National Lab for useful discussions on many aspects of this research.  

\bibliographystyle{apsrev}
\bibliography{strangelets}

\begin{thebibliography}{64}
\expandafter\ifx\csname natexlab\endcsname\relax\def\natexlab#1{#1}\fi
\expandafter\ifx\csname bibnamefont\endcsname\relax
  \def\bibnamefont#1{#1}\fi
\expandafter\ifx\csname bibfnamefont\endcsname\relax
  \def\bibfnamefont#1{#1}\fi
\expandafter\ifx\csname citenamefont\endcsname\relax
  \def\citenamefont#1{#1}\fi
\expandafter\ifx\csname url\endcsname\relax
  \def\url#1{\texttt{#1}}\fi
\expandafter\ifx\csname urlprefix\endcsname\relax\def\urlprefix{URL }\fi
\providecommand{\bibinfo}[2]{#2}
\providecommand{\eprint}[2][]{\url{#2}}

\bibitem[{\citenamefont{Witten}(1984)}]{witten}
\bibinfo{author}{\bibfnamefont{E.}~\bibnamefont{Witten}},
  \bibinfo{journal}{Phys. Rev.} \textbf{\bibinfo{volume}{D30}},
  \bibinfo{pages}{272} (\bibinfo{year}{1984}).

\bibitem[{\citenamefont{Farhi and Jaffe}(1984)}]{farhi}
\bibinfo{author}{\bibfnamefont{E.}~\bibnamefont{Farhi}} \bibnamefont{and}
  \bibinfo{author}{\bibfnamefont{R.~L.} \bibnamefont{Jaffe}},
  \bibinfo{journal}{Phys. Rev.} \textbf{\bibinfo{volume}{D30}},
  \bibinfo{pages}{2379} (\bibinfo{year}{1984}).

\bibitem[{\citenamefont{Madsen}(2001)}]{madsen:CFL}
\bibinfo{author}{\bibfnamefont{J.}~\bibnamefont{Madsen}},
  \bibinfo{journal}{Phys. Rev. Lett.} \textbf{\bibinfo{volume}{87}},
  \bibinfo{pages}{172003} (\bibinfo{year}{2001}), \eprint{hep-ph/0108036}.

\bibitem[{\citenamefont{Madsen}(1998)}]{madsen:star}
\bibinfo{author}{\bibfnamefont{J.}~\bibnamefont{Madsen}},
  \bibinfo{journal}{Phys. Rev. Lett.} \textbf{\bibinfo{volume}{81}},
  \bibinfo{pages}{3311} (\bibinfo{year}{1998}), \eprint{astro-ph/9806032}.

\bibitem[{\citenamefont{Lee et~al.}(2001)\citenamefont{Lee, Kluzniak, and
  Nix}}]{lee02}
\bibinfo{author}{\bibfnamefont{W.~H.} \bibnamefont{Lee}},
  \bibinfo{author}{\bibfnamefont{W.}~\bibnamefont{Kluzniak}}, \bibnamefont{and}
  \bibinfo{author}{\bibfnamefont{J.}~\bibnamefont{Nix}}, \bibinfo{journal}{Acta
  Astron.} \textbf{\bibinfo{volume}{51}}, \bibinfo{pages}{331}
  (\bibinfo{year}{2001}), \eprint{astro-ph/0201114}.

\bibitem[{\citenamefont{Caldwell and Friedman}(1991)}]{caldwell91}
\bibinfo{author}{\bibfnamefont{R.~R.} \bibnamefont{Caldwell}} \bibnamefont{and}
  \bibinfo{author}{\bibfnamefont{J.~R.} \bibnamefont{Friedman}},
  \bibinfo{journal}{Phys. Rev. Lett. B} \textbf{\bibinfo{volume}{264}},
  \bibinfo{pages}{143} (\bibinfo{year}{1991}).

\bibitem[{\citenamefont{Madsen}(2005)}]{madsen:strangelets}
\bibinfo{author}{\bibfnamefont{J.}~\bibnamefont{Madsen}},
  \bibinfo{journal}{Phys. Rev.} \textbf{\bibinfo{volume}{D71}},
  \bibinfo{pages}{014026} (\bibinfo{year}{2005}), \eprint{astro-ph/0411538}.

\bibitem[{\citenamefont{Hemmick et~al.}(1987)}]{hemmick}
\bibinfo{author}{\bibfnamefont{T.~K.} \bibnamefont{Hemmick}}
  \bibnamefont{et~al.}, \bibinfo{journal}{Nucl. Inst. Meth.}
  \textbf{\bibinfo{volume}{B29}} (\bibinfo{year}{1987}).

\bibitem[{\citenamefont{Klingenberg}(2001)}]{Klingenberg}
\bibinfo{author}{\bibfnamefont{R.}~\bibnamefont{Klingenberg}},
  \bibinfo{journal}{J. Phys.} \textbf{\bibinfo{volume}{G27}},
  \bibinfo{pages}{475} (\bibinfo{year}{2001}).

\bibitem[{\citenamefont{Feiveson}(2002)}]{feiveson}
\bibinfo{author}{\bibfnamefont{L.}~\bibnamefont{Feiveson}}, Master's thesis,
  \bibinfo{school}{Yale University} (\bibinfo{year}{2002}).

\bibitem[{\citenamefont{Mueller et~al.}(2004)}]{mueller}
\bibinfo{author}{\bibfnamefont{P.}~\bibnamefont{Mueller}} \bibnamefont{et~al.},
  \bibinfo{journal}{Phys. Rev. Lett.} \textbf{\bibinfo{volume}{92}},
  \bibinfo{pages}{022501} (\bibinfo{year}{2004}), \eprint{nucl-ex/0302025}.

\bibitem[{\citenamefont{Stormer}(1955)}]{stormer2}
\bibinfo{author}{\bibfnamefont{C.}~\bibnamefont{Stormer}},
  \emph{\bibinfo{title}{The Polar Aurora}} (\bibinfo{publisher}{Oxford
  University Press, London}, \bibinfo{year}{1955}).

\bibitem[{\citenamefont{Taylor}(1975)}]{taylor75}
\bibinfo{author}{\bibfnamefont{S.~R.} \bibnamefont{Taylor}},
  \emph{\bibinfo{title}{Lunar science: a post-Apollo view}}
  (\bibinfo{publisher}{Pergamon Press, New York}, \bibinfo{year}{1975}).

\bibitem[{\citenamefont{Eberhart et~al.}(1969)}]{eberhart}
\bibinfo{author}{\bibfnamefont{P.}~\bibnamefont{Eberhart}}
  \bibnamefont{et~al.}, in \emph{\bibinfo{booktitle}{Proceedings of the Apollo
  11 Lunar Science Congress}} (\bibinfo{year}{1969}).

\bibitem[{\citenamefont{Begemann et~al.}(1969)\citenamefont{Begemann, Vilcsek,
  and Wanke}}]{begemann69}
\bibinfo{author}{\bibfnamefont{R.}~\bibnamefont{Begemann},
  \bibfnamefont{F.and~Reider}},
  \bibinfo{author}{\bibfnamefont{E.}~\bibnamefont{Vilcsek}}, \bibnamefont{and}
  \bibinfo{author}{\bibfnamefont{H.}~\bibnamefont{Wanke}}, in
  \emph{\bibinfo{booktitle}{Meteorite Research}}, edited by
  \bibinfo{editor}{\bibfnamefont{P.~M.} \bibnamefont{Millman}}
  (\bibinfo{publisher}{Reidel Publishing Company, Dordrecht},
  \bibinfo{year}{1969}), pp. \bibinfo{pages}{267--274}.

\bibitem[{\citenamefont{Anders}(1962)}]{anders62}
\bibinfo{author}{\bibfnamefont{E.}~\bibnamefont{Anders}},
  \bibinfo{journal}{Rev. Mod. Phys.} \textbf{\bibinfo{volume}{34}},
  \bibinfo{pages}{287} (\bibinfo{year}{1962}).

\bibitem[{\citenamefont{Mohrig}(2004)}]{mohrig}
\bibinfo{author}{\bibfnamefont{D.}~\bibnamefont{Mohrig}},
  \bibinfo{howpublished}{private communication} (\bibinfo{year}{2004}).

\bibitem[{\citenamefont{Salby}(1996)}]{salby}
\bibinfo{author}{\bibfnamefont{M.~L.} \bibnamefont{Salby}},
  \emph{\bibinfo{title}{Fundamentals of Atmospheric Physics}}
  (\bibinfo{publisher}{Academic Press}, \bibinfo{year}{1996}).

\bibitem[{ICA(1958)}]{ICAO}
\emph{\bibinfo{title}{U.S. Extension to the ICAO Standard Atmosphere}},
  \bibinfo{organization}{U.S. Government Printing Office},
  \bibinfo{address}{Washington, D.C.} (\bibinfo{year}{1958}).

\bibitem[{\citenamefont{Craig}(1965)}]{craig}
\bibinfo{author}{\bibfnamefont{R.~A.} \bibnamefont{Craig}},
  \emph{\bibinfo{title}{The Upper Atmosphere: Meteorology and Physics}}
  (\bibinfo{publisher}{Academic Press, NY}, \bibinfo{year}{1965}).

\bibitem[{\citenamefont{Jaffe}(1992)}]{jaffe92}
\bibinfo{author}{\bibfnamefont{D.}~\bibnamefont{Jaffe}}, in
  \emph{\bibinfo{booktitle}{Global Biogeochemical Cycles}}, edited by
  \bibinfo{editor}{\bibfnamefont{S.}~\bibnamefont{Butcher}},
  \bibinfo{editor}{\bibfnamefont{R.~J.} \bibnamefont{Charlson}},
  \bibinfo{editor}{\bibfnamefont{G.~H.} \bibnamefont{Orians}},
  \bibnamefont{and} \bibinfo{editor}{\bibfnamefont{G.~V.} \bibnamefont{Wolfe}}
  (\bibinfo{publisher}{Academic Press, San Diego}, \bibinfo{year}{1992}), pp.
  \bibinfo{pages}{263--284}.

\bibitem[{\citenamefont{Keeling et~al.}(1993)\citenamefont{Keeling, Najjar,
  Bender, and Tans}}]{keeling93}
\bibinfo{author}{\bibfnamefont{R.~F.} \bibnamefont{Keeling}},
  \bibinfo{author}{\bibfnamefont{R.}~\bibnamefont{Najjar}},
  \bibinfo{author}{\bibfnamefont{M.}~\bibnamefont{Bender}}, \bibnamefont{and}
  \bibinfo{author}{\bibfnamefont{P.}~\bibnamefont{Tans}},
  \bibinfo{journal}{Global Biogeochemical Cycles} \textbf{\bibinfo{volume}{7}},
  \bibinfo{pages}{679} (\bibinfo{year}{1993}).

\bibitem[{\citenamefont{Hunter}(1980)}]{hunter80}
\bibinfo{author}{\bibfnamefont{D.}~\bibnamefont{Hunter}}, \bibinfo{journal}{J.
  Atmos. Sci.} \textbf{\bibinfo{volume}{37}}, \bibinfo{pages}{1342}
  (\bibinfo{year}{1980}).

\bibitem[{\citenamefont{Cziczo et~al.}(2001)\citenamefont{Cziczo, Thomson, and
  Murphy}}]{cziczo}
\bibinfo{author}{\bibfnamefont{D.~J.} \bibnamefont{Cziczo}},
  \bibinfo{author}{\bibfnamefont{D.~S.} \bibnamefont{Thomson}},
  \bibnamefont{and} \bibinfo{author}{\bibfnamefont{D.~M.}
  \bibnamefont{Murphy}}, \bibinfo{journal}{Science}
  \textbf{\bibinfo{volume}{291}}, \bibinfo{pages}{1772} (\bibinfo{year}{2001}).

\bibitem[{\citenamefont{Bigeleisen}(1975)}]{bigeleisen_book}
\bibinfo{author}{\bibfnamefont{J.}~\bibnamefont{Bigeleisen}}, in
  \emph{\bibinfo{booktitle}{Isotopes and Chemical Principles}}, edited by
  \bibinfo{editor}{\bibfnamefont{P.~A.} \bibnamefont{Rock}}
  (\bibinfo{publisher}{American Chemical Society}, \bibinfo{year}{1975}), pp.
  \bibinfo{pages}{1--28}.

\bibitem[{\citenamefont{Br\"ugger et~al.}(1989)\citenamefont{Br\"ugger,
  L\"utzenkirchen, Polikanov, Herrmann, Overbeck, Trautmann, Breskin, Chechik,
  Fraenkel, and Smilansky}}]{brugger}
\bibinfo{author}{\bibfnamefont{M.}~\bibnamefont{Br\"ugger}},
  \bibinfo{author}{\bibfnamefont{K.}~\bibnamefont{L\"utzenkirchen}},
  \bibinfo{author}{\bibfnamefont{S.}~\bibnamefont{Polikanov}},
  \bibinfo{author}{\bibfnamefont{G.}~\bibnamefont{Herrmann}},
  \bibinfo{author}{\bibfnamefont{M.}~\bibnamefont{Overbeck}},
  \bibinfo{author}{\bibfnamefont{N.}~\bibnamefont{Trautmann}},
  \bibinfo{author}{\bibfnamefont{A.}~\bibnamefont{Breskin}},
  \bibinfo{author}{\bibfnamefont{R.}~\bibnamefont{Chechik}},
  \bibinfo{author}{\bibfnamefont{Z.}~\bibnamefont{Fraenkel}}, \bibnamefont{and}
  \bibinfo{author}{\bibfnamefont{U.}~\bibnamefont{Smilansky}},
  \bibinfo{journal}{Nature} \textbf{\bibinfo{volume}{337}},
  \bibinfo{pages}{434} (\bibinfo{year}{1989}).

\bibitem[{\citenamefont{Perillo~Isaac et~al.}(1998)}]{perillo}
\bibinfo{author}{\bibfnamefont{M.~C.} \bibnamefont{Perillo~Isaac}}
  \bibnamefont{et~al.}, \bibinfo{journal}{Phys. Rev. Lett.}
  \textbf{\bibinfo{volume}{81}}, \bibinfo{pages}{2416} (\bibinfo{year}{1998}),
  \eprint{astro-ph/9806147}.

\bibitem[{\citenamefont{Lowder}(1991)}]{lowder91}
\bibinfo{author}{\bibfnamefont{D.}~\bibnamefont{Lowder}}, in
  \emph{\bibinfo{booktitle}{International Workshop on Strange Quark Matter
  Physics and Astrophysics}}, edited by
  \bibinfo{editor}{\bibfnamefont{J.}~\bibnamefont{Madsen}} \bibnamefont{and}
  \bibinfo{editor}{\bibfnamefont{P.}~\bibnamefont{Haemel}}
  (\bibinfo{publisher}{North-Holland}, \bibinfo{address}{University of Aarhus,
  Denmark}, \bibinfo{year}{1991}).

\bibitem[{\citenamefont{Takahashi and Boyd}(1988)}]{takahashi}
\bibinfo{author}{\bibfnamefont{K.}~\bibnamefont{Takahashi}} \bibnamefont{and}
  \bibinfo{author}{\bibfnamefont{R.}~\bibnamefont{Boyd}},
  \bibinfo{journal}{Astrophys. J} \textbf{\bibinfo{volume}{327}},
  \bibinfo{pages}{1009} (\bibinfo{year}{1988}).

\bibitem[{\citenamefont{Vandegriff et~al.}(1996)\citenamefont{Vandegriff,
  Raimann, Boyd, Caffee, and Ruiz}}]{vandegriff}
\bibinfo{author}{\bibfnamefont{J.}~\bibnamefont{Vandegriff}},
  \bibinfo{author}{\bibfnamefont{G.}~\bibnamefont{Raimann}},
  \bibinfo{author}{\bibfnamefont{R.~N.} \bibnamefont{Boyd}},
  \bibinfo{author}{\bibfnamefont{M.}~\bibnamefont{Caffee}}, \bibnamefont{and}
  \bibinfo{author}{\bibfnamefont{B.}~\bibnamefont{Ruiz}},
  \bibinfo{journal}{Phys. Lett. B} \textbf{\bibinfo{volume}{365}},
  \bibinfo{pages}{418} (\bibinfo{year}{1996}).

\bibitem[{\citenamefont{Holt et~al.}(1976)}]{holt}
\bibinfo{author}{\bibfnamefont{R.}~\bibnamefont{Holt}} \bibnamefont{et~al.},
  \bibinfo{journal}{Phys. Rev. Letters} \textbf{\bibinfo{volume}{36}},
  \bibinfo{pages}{183} (\bibinfo{year}{1976}).

\bibitem[{\citenamefont{Stein and Hohorst}(1982)}]{stein1982}
\bibinfo{author}{\bibfnamefont{L.}~\bibnamefont{Stein}} \bibnamefont{and}
  \bibinfo{author}{\bibfnamefont{F.~A.} \bibnamefont{Hohorst}},
  \bibinfo{journal}{Environ. Sci. Technol.} \textbf{\bibinfo{volume}{16}},
  \bibinfo{pages}{419} (\bibinfo{year}{1982}).

\bibitem[{\citenamefont{Berger and Jaffe}(1987)}]{jaffe_radioactivity}
\bibinfo{author}{\bibfnamefont{M.~S.} \bibnamefont{Berger}} \bibnamefont{and}
  \bibinfo{author}{\bibfnamefont{R.~L.} \bibnamefont{Jaffe}},
  \bibinfo{journal}{Phys. Rev.} \textbf{\bibinfo{volume}{C35}},
  \bibinfo{pages}{213} (\bibinfo{year}{1987}).

\bibitem[{\citenamefont{{Sigma Aldrich Co.}}(2004)}]{sigmaaldrich}
\bibinfo{author}{\bibnamefont{{Sigma Aldrich Co.}}},
  \bibinfo{howpublished}{private communication} (\bibinfo{year}{2004}).

\bibitem[{\citenamefont{Arnold et~al.}(1998)\citenamefont{Arnold, Curtius,
  Spreng, and Deshler}}]{arnold98}
\bibinfo{author}{\bibfnamefont{F.}~\bibnamefont{Arnold}},
  \bibinfo{author}{\bibfnamefont{J.}~\bibnamefont{Curtius}},
  \bibinfo{author}{\bibfnamefont{S.}~\bibnamefont{Spreng}}, \bibnamefont{and}
  \bibinfo{author}{\bibfnamefont{T.}~\bibnamefont{Deshler}},
  \bibinfo{journal}{J. Atmos. Chem} \textbf{\bibinfo{volume}{30}},
  \bibinfo{pages}{3} (\bibinfo{year}{1998}).

\bibitem[{\citenamefont{van Hook}(1975)}]{van_hook_book}
\bibinfo{author}{\bibfnamefont{W.~A.} \bibnamefont{van Hook}}, in
  \emph{\bibinfo{booktitle}{Isotopes and Chemical Principles}}, edited by
  \bibinfo{editor}{\bibfnamefont{P.~A.} \bibnamefont{Rock}}
  (\bibinfo{publisher}{American Chemical Society}, \bibinfo{year}{1975}), pp.
  \bibinfo{pages}{101--130}.

\bibitem[{\citenamefont{Keesom et~al.}(1934)\citenamefont{Keesom, Van~Dijk, and
  Haantjes}}]{keesom}
\bibinfo{author}{\bibfnamefont{W.}~\bibnamefont{Keesom}},
  \bibinfo{author}{\bibfnamefont{H.}~\bibnamefont{Van~Dijk}}, \bibnamefont{and}
  \bibinfo{author}{\bibfnamefont{J.}~\bibnamefont{Haantjes}},
  \bibinfo{journal}{Physica} \textbf{\bibinfo{volume}{1}},
  \bibinfo{pages}{1109} (\bibinfo{year}{1934}).

\bibitem[{\citenamefont{Clusius and Mayer}(1953)}]{clusius}
\bibinfo{author}{\bibfnamefont{K.}~\bibnamefont{Clusius}} \bibnamefont{and}
  \bibinfo{author}{\bibfnamefont{M.~G.} \bibnamefont{Mayer}},
  \bibinfo{journal}{Helv. Chim. Acta} \textbf{\bibinfo{volume}{36}}
  (\bibinfo{year}{1953}).

\bibitem[{\citenamefont{Canongia~Lopes
  et~al.}(2002)\citenamefont{Canongia~Lopes, Rebelo, and Bigeleisen}}]{kr_vpie}
\bibinfo{author}{\bibfnamefont{J.~N.} \bibnamefont{Canongia~Lopes}},
  \bibinfo{author}{\bibfnamefont{L.~P.~N.} \bibnamefont{Rebelo}},
  \bibnamefont{and}
  \bibinfo{author}{\bibfnamefont{J.}~\bibnamefont{Bigeleisen}},
  \bibinfo{journal}{J. Chem. Phys.} \textbf{\bibinfo{volume}{117}},
  \bibinfo{pages}{8836} (\bibinfo{year}{2002}).

\bibitem[{\citenamefont{Jancso and Van~Hook}(1974)}]{jancso}
\bibinfo{author}{\bibfnamefont{G.}~\bibnamefont{Jancso}} \bibnamefont{and}
  \bibinfo{author}{\bibfnamefont{W.~A.} \bibnamefont{Van~Hook}},
  \bibinfo{journal}{Chem. Rev.} \textbf{\bibinfo{volume}{74}},
  \bibinfo{pages}{689} (\bibinfo{year}{1974}).

\bibitem[{\citenamefont{London}(1957)}]{london1958}
\bibinfo{author}{\bibfnamefont{H.}~\bibnamefont{London}}, in
  \emph{\bibinfo{booktitle}{Proceedings of the International Symposium on
  Isotope Separation}}, edited by
  \bibinfo{editor}{\bibfnamefont{J.}~\bibnamefont{Kistemaker}}
  (\bibinfo{publisher}{North-Holland}, \bibinfo{year}{1957}).

\bibitem[{\citenamefont{Vasaru et~al.}(1969)\citenamefont{Vasaru, M\"uller,
  Reinhold, and Fodor}}]{vasaru}
\bibinfo{author}{\bibfnamefont{G.}~\bibnamefont{Vasaru}},
  \bibinfo{author}{\bibfnamefont{G.}~\bibnamefont{M\"uller}},
  \bibinfo{author}{\bibfnamefont{G.}~\bibnamefont{Reinhold}}, \bibnamefont{and}
  \bibinfo{author}{\bibfnamefont{T.}~\bibnamefont{Fodor}},
  \emph{\bibinfo{title}{The Thermal Diffusion Column}} (\bibinfo{publisher}{Veb
  Deutscher Verlag der Wissenshaften}, \bibinfo{year}{1969}).

\bibitem[{\citenamefont{London}(1961)}]{london}
\bibinfo{editor}{\bibfnamefont{H.}~\bibnamefont{London}}, ed.,
  \emph{\bibinfo{title}{Separation of Isotopes}} (\bibinfo{publisher}{George
  Newnes Limited}, \bibinfo{year}{1961}).

\bibitem[{\citenamefont{Spindel}(1975)}]{spindel_book}
\bibinfo{author}{\bibfnamefont{W.}~\bibnamefont{Spindel}}, in
  \emph{\bibinfo{booktitle}{Isotopes and Chemical Principles}}, edited by
  \bibinfo{editor}{\bibfnamefont{P.~A.} \bibnamefont{Rock}}
  (\bibinfo{publisher}{American Chemical Society}, \bibinfo{year}{1975}), pp.
  \bibinfo{pages}{1--28}.

\bibitem[{\citenamefont{Cohen and Cohen}(1983)}]{cohen}
\bibinfo{author}{\bibfnamefont{B.}~\bibnamefont{Cohen}} \bibnamefont{and}
  \bibinfo{author}{\bibfnamefont{E.}~\bibnamefont{Cohen}},
  \bibinfo{journal}{Health Physics} \textbf{\bibinfo{volume}{45}}
  (\bibinfo{year}{1983}).

\bibitem[{\citenamefont{Fujii}()}]{fujii}
\bibinfo{author}{\bibfnamefont{Y.}~\bibnamefont{Fujii}},
  \emph{\bibinfo{title}{various works}},
  \bibinfo{howpublished}{http://www.nr.titech.ac.jp/$\sim$yfujii/result.htm}.

\bibitem[{\citenamefont{Bigeleisen and Mayer}(1947)}]{bigeleisen}
\bibinfo{author}{\bibfnamefont{J.}~\bibnamefont{Bigeleisen}} \bibnamefont{and}
  \bibinfo{author}{\bibfnamefont{M.~G.} \bibnamefont{Mayer}},
  \bibinfo{journal}{J. Chem. Phys.} \textbf{\bibinfo{volume}{15}},
  \bibinfo{pages}{261} (\bibinfo{year}{1947}).

\bibitem[{\citenamefont{Ban et~al.}(2002)\citenamefont{Ban, Nomura, and
  Fujii}}]{li_chrom}
\bibinfo{author}{\bibfnamefont{Y.}~\bibnamefont{Ban}},
  \bibinfo{author}{\bibfnamefont{M.}~\bibnamefont{Nomura}}, \bibnamefont{and}
  \bibinfo{author}{\bibfnamefont{Y.}~\bibnamefont{Fujii}}, \bibinfo{journal}{J.
  Nucl. Sci. Tech.} \textbf{\bibinfo{volume}{39}}, \bibinfo{pages}{279}
  (\bibinfo{year}{2002}).

\bibitem[{\citenamefont{Kim}(2001)}]{mg_chrom}
\bibinfo{author}{\bibfnamefont{D.~W.} \bibnamefont{Kim}}, \bibinfo{journal}{J.
  Nucl. Sci. Tech.} \textbf{\bibinfo{volume}{38}}, \bibinfo{pages}{780}
  (\bibinfo{year}{2001}).

\bibitem[{\citenamefont{Zhang et~al.}(2003)\citenamefont{Zhang, Nomura, Aida,
  and Fujii}}]{zhang_v}
\bibinfo{author}{\bibfnamefont{Y.~H.} \bibnamefont{Zhang}},
  \bibinfo{author}{\bibfnamefont{M.}~\bibnamefont{Nomura}},
  \bibinfo{author}{\bibfnamefont{M.}~\bibnamefont{Aida}}, \bibnamefont{and}
  \bibinfo{author}{\bibfnamefont{Y.}~\bibnamefont{Fujii}}, \bibinfo{journal}{J.
  Chromatography A} \textbf{\bibinfo{volume}{989}}, \bibinfo{pages}{175}
  (\bibinfo{year}{2003}).

\bibitem[{\citenamefont{Ban and Nomura}(2002)}]{zn_chrom}
\bibinfo{author}{\bibfnamefont{Y.}~\bibnamefont{Ban}} \bibnamefont{and}
  \bibinfo{author}{\bibfnamefont{Y.}~\bibnamefont{Nomura},
  \bibfnamefont{M.and~Fujii}}, \bibinfo{journal}{J. Nucl. Sci. and Tech.}
  \textbf{\bibinfo{volume}{39}}, \bibinfo{pages}{156} (\bibinfo{year}{2002}).

\bibitem[{\citenamefont{Fujii et~al.}(2002)}]{ba_chrom}
\bibinfo{author}{\bibfnamefont{T.}~\bibnamefont{Fujii}} \bibnamefont{et~al.},
  \bibinfo{journal}{J. Nucl. Sci. Tech.} \textbf{\bibinfo{volume}{39}},
  \bibinfo{pages}{447} (\bibinfo{year}{2002}).

\bibitem[{\citenamefont{Ismail et~al.}(2002)}]{u_chrom}
\bibinfo{author}{\bibfnamefont{I.~M.} \bibnamefont{Ismail}}
  \bibnamefont{et~al.}, \bibinfo{journal}{Z. Naturforsch.}
  \textbf{\bibinfo{volume}{57a}}, \bibinfo{pages}{247} (\bibinfo{year}{2002}).

\bibitem[{\citenamefont{Sandweiss et~al.}()}]{sandweiss}
\bibinfo{author}{\bibfnamefont{J.}~\bibnamefont{Sandweiss}}
  \bibnamefont{et~al.}, \bibinfo{note}{unpublished research, Wright Nuclear
  Structure Lab}.

\bibitem[{\citenamefont{Bailey}(2000)}]{atta_nimb}
\bibinfo{author}{\bibfnamefont{K.}~\bibnamefont{Bailey}},
  \bibinfo{journal}{Nucl. Instrum. Meth.} \textbf{\bibinfo{volume}{B172}},
  \bibinfo{pages}{224} (\bibinfo{year}{2000}).

\bibitem[{\citenamefont{Jarvis et~al.}(1992)\citenamefont{Jarvis, Gray, and
  Houk}}]{jarvis92}
\bibinfo{author}{\bibfnamefont{K.~E.} \bibnamefont{Jarvis}},
  \bibinfo{author}{\bibfnamefont{A.~L.} \bibnamefont{Gray}}, \bibnamefont{and}
  \bibinfo{author}{\bibfnamefont{R.~S.} \bibnamefont{Houk}},
  \emph{\bibinfo{title}{Handbook of Inductively Coupled Mass Spectrometry}}
  (\bibinfo{publisher}{Chapman and Hall}, \bibinfo{year}{1992}).

\bibitem[{\citenamefont{Marshall}(1985)}]{marshall85}
\bibinfo{author}{\bibfnamefont{A.}~\bibnamefont{Marshall}},
  \bibinfo{journal}{Acc. Chem. Res.} \textbf{\bibinfo{volume}{18}},
  \bibinfo{pages}{316} (\bibinfo{year}{1985}).

\bibitem[{\citenamefont{Belov et~al.}(2000)\citenamefont{Belov, Gorshkov,
  Udseth, Anderson, and Smith}}]{belov}
\bibinfo{author}{\bibfnamefont{M.}~\bibnamefont{Belov}},
  \bibinfo{author}{\bibfnamefont{M.}~\bibnamefont{Gorshkov}},
  \bibinfo{author}{\bibfnamefont{H.}~\bibnamefont{Udseth}},
  \bibinfo{author}{\bibfnamefont{G.}~\bibnamefont{Anderson}}, \bibnamefont{and}
  \bibinfo{author}{\bibfnamefont{R.}~\bibnamefont{Smith}},
  \bibinfo{journal}{Anal Chem} \textbf{\bibinfo{volume}{72}},
  \bibinfo{pages}{2271} (\bibinfo{year}{2000}).

\bibitem[{\citenamefont{Tisato et~al.}(2004)}]{tisato04}
\bibinfo{author}{\bibfnamefont{F.}~\bibnamefont{Tisato}} \bibnamefont{et~al.},
  \bibinfo{journal}{Mass Spectrometry Reviews} \textbf{\bibinfo{volume}{23}},
  \bibinfo{pages}{309} (\bibinfo{year}{2004}).

\bibitem[{\citenamefont{Wendt et~al.}(2000)}]{wendt00}
\bibinfo{author}{\bibfnamefont{K.}~\bibnamefont{Wendt}} \bibnamefont{et~al.},
  \bibinfo{journal}{Nucl. Inst. Meth.} \textbf{\bibinfo{volume}{B172}},
  \bibinfo{pages}{162} (\bibinfo{year}{2000}).

\bibitem[{\citenamefont{Pellin et~al.}(2001)\citenamefont{Pellin, Calaway, and
  Veryovkin}}]{sarisa_book}
\bibinfo{author}{\bibfnamefont{M.~J.} \bibnamefont{Pellin}},
  \bibinfo{author}{\bibfnamefont{W.~F.} \bibnamefont{Calaway}},
  \bibnamefont{and}
  \bibinfo{author}{\bibfnamefont{I.}~\bibnamefont{Veryovkin}}, in
  \emph{\bibinfo{booktitle}{ToFSIMS: Surface Analysis by Mass Spectrometry}},
  edited by \bibinfo{editor}{\bibfnamefont{J.}~\bibnamefont{Vickerman}}
  \bibnamefont{and} \bibinfo{editor}{\bibfnamefont{D.}~\bibnamefont{Briggs}}
  (\bibinfo{publisher}{IM Publications}, \bibinfo{year}{2001}), pp.
  \bibinfo{pages}{375--415}.

\bibitem[{\citenamefont{Vitousek et~al.}(1997)}]{vitousek}
\bibinfo{author}{\bibfnamefont{P.~M.} \bibnamefont{Vitousek}}
  \bibnamefont{et~al.}, \bibinfo{journal}{Issues in Ecology}
  \textbf{\bibinfo{volume}{1}} (\bibinfo{year}{1997}).

\bibitem[{\citenamefont{Lehninger et~al.}(1993)\citenamefont{Lehninger, Nelson,
  and Cox}}]{biochemistry}
\bibinfo{author}{\bibfnamefont{A.~L.} \bibnamefont{Lehninger}},
  \bibinfo{author}{\bibfnamefont{D.~L.} \bibnamefont{Nelson}},
  \bibnamefont{and} \bibinfo{author}{\bibfnamefont{M.~M.} \bibnamefont{Cox}},
  \emph{\bibinfo{title}{Principles of Biochemistry}}
  (\bibinfo{publisher}{Worth, New York}, \bibinfo{year}{1993}),
  \bibinfo{edition}{2nd} ed.

\bibitem[{\citenamefont{Ishida}(2002)}]{ishida}
\bibinfo{author}{\bibfnamefont{T.}~\bibnamefont{Ishida}}, \bibinfo{journal}{J.
  Nucl. Sci. Tech.} \textbf{\bibinfo{volume}{39}}, \bibinfo{pages}{407}
  (\bibinfo{year}{2002}).

\end{thebibliography}

\end{document}